\documentclass[12pt,cite,epsf,epsfig]{article}
\usepackage{epsfig}
\usepackage{amsmath}
\usepackage{cite}

\begin{document}
\title{Neutrino Mass Matrices with Two Vanishing Elements/Cofactors}
\author{S. Dev$^{a, }$\thanks{sdev@iucaa.ernet.in}, Lal Singh$^{b, }$\thanks{lalsingh96@yahoo.com}, and Desh Raj$^{b, }$\thanks{raj.physics88@gmail.com}}
\date{$^a$\textit{Department of Physics, School of Sciences, HNBG Central University, Srinagar, Uttarakhand 246174, India}\\
\smallskip
$^b$\textit{Department of Physics, Himachal Pradesh University, Shimla 171005, India}
}

\maketitle
\begin{abstract}
We study the phenomenological implications of the recent neutrino data for class B of two texture zeros and two vanishing cofactors for Majorana neutrinos in the flavor basis. We find that classes $B_{1}$($B_2$) of two texture zeros and classes $B_5$($B_6$) of two vanishing cofactors have similar predictions for neutrino oscillation parameters for the same mass hierarchy. Similar predictions for classes $B_3$($B_4$) of two texture zeros and classes $B_3$($B_4$) of two vanishing cofactors are expected. However, a preference for a shift in the quadrant of the Dirac-type CP violating phase($\delta$) in contrast to the earlier analysis has been predicted for a relatively large value of the reactor neutrino mixing angle($\theta_{13}$) for class B of two texture zeros and two vanishing cofactors for an inverted mass spectrum. No such shift in the quadrant of $\delta$ has been found for the normal mass spectrum.
\end{abstract}

\section{Introduction}
There has been a significant progress in precise determination of neutrino masses and mixings during the recent past. Recently, a non-zero and relatively large value of the last unknown mixing angle ($\theta_{13}$) has been confirmed by a number of neutrino oscillation experiments\cite{t2k, minos, dchooz, dayabay, reno} with a best fit around $9^{\circ}$. 
Relatively large value of $\theta_{13}$ has brightened the prospects for exploring the Dirac-type CP-violating phase $\delta$ in the lepton sector, determining the octant of atmospheric mixing angle ($\theta_{23}$) and identifying neutrino mass hierarchy. However, the currently available data on neutrino masses and mixings  is not enough to determine the neutrino mass matrix($M_{\nu}$) unambiguously and additional theoretical inputs such as zero textures\cite{zero,tz,zerotexture}, vanishing minors\cite{Lashin,zerominor,tzm}, hybrid textures\cite{hybrid} and  equalities between elements \cite{tec} amongst others which restrict the structure of $M_{\nu}$ become necessary for further progress.

Neutrino mass matrices with two texture zeros or two vanishing cofactors are the most extensively studied texture structures. In the flavour basis, with the charged lepton mass matrix diagonal, there are 15 possible cases of two texture zeros (TZ) in the effective neutrino mass matrix $M_\nu$ of which only 7 are compatible with the current neutrino oscillation data \cite{valledata}. Similarly, in case of two vanishing cofactors (TC) only 7 out of the possible 15 texture structures survive the current neutrino experimental constraints. The viable texture structures in case of TZ and TC are listed in Table 1. $A_1,A_2,B_3~\textrm{and}~B_4$ texture structures of TZ in the nomenclature of Ref. \cite{zero} are equivalent to $A_1,A_2,B_3~\textrm{and}~B_4$ class of TC, respectively in the nomenclature of Ref. \cite{Lashin}. $B_1,B_2~\textrm{and}~C$ of texture zeros do not appear as two zero cofactors and $B_5,B_6~\textrm{and}~D$ of TC provide non-trivial vanishing cofactors when confronted with the neutrino oscillation data.
\begin{table}[h]
\begin{small}
\begin{center}
\begin{tabular}{cc| cc}\hline \hline
 Class  & Texture Zeros & Class  & Zero Cofactors \\ 
\hline $A_1$ & $M_{11}$, $M_{12}$ & $A_1$ & $C_{33}$, $C_{32}$ \\ 
 $A_2$ & $M_{11}$, $M_{13}$ & $A_2$ & $C_{22}$, $C_{32}$ \\ 
 $B_1$ & $M_{13}$, $M_{22}$ & $B_3$ & $C_{33}$, $C_{13}$ \\ 
 $B_2$ & $M_{12}$, $M_{33}$ & $B_4$ & $C_{22}$, $C_{21}$\\ 
 $B_3$ & $M_{12}$, $M_{22}$ & $B_5$ & $C_{33}$, $C_{12}$ \\ 
 $B_4$ & $M_{13}$, $M_{33}$ & $B_6$ & $C_{22}$, $C_{13}$\\ 
 $C$   & $M_{22}$, $M_{33}$ & $D$   & $C_{33}$, $C_{22}$\\ 
 \hline \hline
\end{tabular}
\caption{Experimentally viable classes of two texture zeros \cite{zero}(two vanishing cofactors \cite{Lashin}), here $M_{ij}$($C_{ij}$) denotes the texture zero(vanishing cofactors) for the $(ij)^{th}$ element of $M_\nu$.}
\end{center}
\end{small} 
\end{table}

Even before the confirmation of relatively large value of $\theta_{13}$, TZ and TC have been studied comprehensively in the recent past. The signatures for distinguishing the various texture structures TZ and TC on the basis of octant of $\theta_{23}$, quadrant of $\delta$ and mass hierarchy have been given in Ref. \cite{tz,Lashin}. After the confirmation of relatively large value of $\theta_{13}$, these texture structures have been further checked for their viability and predictions for the unknown neutrino parameters like $\delta$, Majorana phases and neutrino mass scale \cite{tznew}. The indistinguishability of TZ and TC has been studied in \cite{Lashin}. Recently, Liao et al. \cite{Liao} have discussed the indistinguishability of $B_1$($B_{2}$) class of TZ and $B_6$($B_5$) class of TC with opposite mass hierarchy, respectively. Their analysis claims similar predictions for the oscillation parameters and Majorana phases which means that if mass hierarchy is not known texture $B_1$($B_2$) of TZ and $B_6$($B_5$) of TC are indistinguishable but the results obtained earlier in \cite{tz,Lashin} seem to distinguish the TZ and TC on the basis of quadrant of $\delta$. The recent global neutrino analyses favour $\delta$ around $3\pi/2$ \cite{1405,1409} which motivates us to re-analyse the TZ and TC for their indistinguishability and predictions for the quadrant of $\delta$ with the recent neutrino experimental data. Texture $B_3$($B_4$) of TZ appears as $B_3$($B_4$) of TC and, therefore, shows identical predictions. Classes ($B_1$,$B_2$) of TZ and classes ($B_5$, $B_6$) of TC are $P_{23}$\footnote{For $P_{23}$ symmetry see equations[18-20]} symmetric texture structures, respectively. Therefore, these texture structures in TZ or in TC differ only in their predictions for the octant of $\theta_{23}$ and quadrant of $\delta$.  A close look at $B_{3}$($B_4$) of TZ and $B_{3}$($B_4$) of TC reveals that zeros and cofactors in these textures are 2-3 interchange symmetric. Similar interchange symmetry is present in $B_1$($B_2$) of TZ and $B_5$($B_6$) of TC structures. Unlike $B_{3}$($B_4$) of TZ and TC having  trivial texture structure, it is interesting to see how the $B_1$($B_2$) of TZ is related to $B_5$($B_6$) of TC for similar mass spectrum.
 
In this work, we present a phenomenological analysis for class B of TZ and TC in the light of the recent experimental data. The identical results for $B_3$($B_4$) of TZ and $B_3$($B_4$) of TC are trivial. However, we observe an interesting pattern of  correlation between neutrino oscillation parameters for $B_1$($B_2$) of TZ and $B_5$($B_6$) of TC. Also, a relatively large value of $\theta_{13}$ has interesting implications for the quadrant of Dirac-type CP-violating phase $\delta$. A particular quadrant for  $\delta$ has been predicted for normal mass spectrum as predicted earlier in Refs. \cite{tz, Lashin}. However, a curious preference for a shift in the quadrant of $\delta$ in contrast to the results reported in Refs. \cite{tz, Lashin} has been predicted for an inverted mass spectrum for texture structures in class B of TZ and TC.

\section{Formalism}
In the flavor basis, where the charged lepton mass matrix $M_l$ is diagonal, the complex symmetric Majorana neutrino mass matrix can be diagonalized by a unitary matrix $V'$ as
\begin{equation}
M_{\nu}= V' M_{\nu}^{diag}V'^{T}
\end{equation}
where $M_{\nu}^{diag}$ = diag$(m_1,m_2,m_3)$. 
The unitary matrix $V'$ can be parametrized as
\begin{equation}
V' = P_lV \ \ \ \textrm{with}\ \ \ \ V = UP_\nu
\end{equation}
where  \cite{foglipdg}
\begin{equation}
U= \left(
\begin{array}{ccc}
c_{12}c_{13} & s_{12}c_{13} & s_{13}e^{-i\delta} \\
-s_{12}c_{23}-c_{12}s_{23}s_{13}e^{i\delta} &
c_{12}c_{23}-s_{12}s_{23}s_{13}e^{i\delta} & s_{23}c_{13} \\
s_{12}s_{23}-c_{12}c_{23}s_{13}e^{i\delta} &
-c_{12}s_{23}-s_{12}c_{23}s_{13}e^{i\delta} & c_{23}c_{13}
\end{array}
\right)
\end{equation} with $s_{ij}=\sin\theta_{ij}$ and $c_{ij}=\cos\theta_{ij}$ and
\begin{small}
\begin{center}
$P_\nu = \left(
\begin{array}{ccc}
1 & 0 & 0 \\ 0 & e^{i\alpha} & 0 \\ 0 & 0 & e^{i(\beta+\delta)}
\end{array}
\right) , \ \ \ 
P_l = \left(
\begin{array}{ccc}
e^{i\varphi_e} & 0 & 0 \\ 0 & e^{i\varphi_\mu} & 0 \\ 0 & 0 & e^{i\varphi_\tau}
\end{array}
\right).$
\end{center}
\end{small}
$P_\nu$ is the diagonal phase matrix containing two Majorana-type CP- violating phases $\alpha$, $\beta$ and one Dirac-type CP-violating phase $\delta$. The phase matrix $P_l$ is physically unobservable and depends on the phase convention. The matrix $V$ is called the neutrino mixing matrix or the Pontecorvo-Maki-Nakagawa-Sakata (PMNS) matrix  \cite{pmns}. Using Eqs.(1) and (2), the effective Majorana neutrino mass matrix can be written as
\begin{equation}
M_{\nu}=P_l U P_\nu M_{\nu}^{diag}P_\nu^{T}U^{T}P_l^T.
\end{equation}
The Dirac-type CP violation in neutrino oscillation experiments can be described through a rephasing invariant quantity, $J_{CP}$ \cite{jarlskog} with $J_{CP}=Im(U_{e1}U_{\mu2}U_{e2}^*U_{\mu1}^*)$. In the above parametrization, $J_{CP}$ is given by
\begin{equation}
J_{CP} = s_{12}s_{23}s_{13}c_{12}c_{23}c_{13}^2 \sin \delta \ .
\end{equation}
\subsection{Two Texture Zeros in $M_\nu$}
The simultaneous existence of two zeros in the neutrino mass matrix at (a,b) and (c,d) positions implies
\begin{align}
& M_{\nu (ab)} = 0~\textrm{and}~M_{\nu (cd)} = 0 \ .
\end{align}
The above complex equations can be written as
\begin{eqnarray}
m_1 A_1 + m_2 A_2 e^{2i\alpha} + m_3 A_3 e^{2i(\beta +\delta)}=0 \ , \\
m_1 B_1 + m_2 B_2 e^{2i\alpha} + m_3 B_3 e^{2i(\beta +\delta)}=0 \ ,
\end{eqnarray}
where
\begin{equation}
A_i = U_{ai}U_{bi} \ \textrm{and} \ B_i = U_{ci}U_{di} \ ,
\end{equation}
with $i = 1, 2, 3$ and a,b,c,d can take values $e,~\mu,~\tau$. These two complex Eqs. (7) and (8) involve nine physical parameters viz. the three neutrino masses ($m_{1}$, $m_{2}$, $m_{3}$), three mixing angles ($\theta _{12}$, $\theta _{23}$, $\theta _{13}$), two Majorana-type CP-violating phases ($\alpha $, $\beta $) and one Dirac-type CP-violating phase ($\delta $). The masses $m_{2}$ and $m_{3}$ can be calculated from the mass-squared differences $\Delta m_{21}^{2}$ and $|\Delta m_{32}^{2}|$ using the relations
\begin{equation}
m_{2}=\sqrt{m_{1}^{2}+\Delta m_{21}^{2}} \ , \ \  m_{3}=\sqrt{m_{2}^{2}+|\Delta m_{32}^{2}|} \ 
\end{equation}
where  $\Delta m_{ij}^{2}=m_i^2-m_j^2$,~ $m_2 < m_3$ for Normal Hierarchy (NH) and $m_2 > m_3$ for Inverted Hierarchy (IH). 
Simultaneously solving Eqs. (7) and (8) for the two mass ratios, we obtain
\begin{small}
\begin{equation}
\frac{m_1}{m_2}e^{-2i\alpha }=\frac{A_2 B_3 - A_3 B_2}{A_3 B_1 - A_1 B_3}
\end{equation}
\end{small}
and
\begin{small}
\begin{equation}
\frac{m_1}{m_3}e^{-2i\beta }=\frac{A_3 B_2 - A_2 B_3 }{A_2 B_1- A_1 B_2}e^{2i\delta } \ .
\end{equation}
\end{small}

The magnitudes of the two mass ratios in Eqs.(11) and (12) are given by
\begin{equation}
\rho=\left|\frac{m_1}{m_3}e^{-2i\beta }\right| ,
\end{equation}
\begin{equation}
\sigma=\left|\frac{m_1}{m_2}e^{-2i\alpha }\right| .
\end{equation}
 while the CP- violating Majorana phases $\alpha$ and $\beta$ are given by
 \begin{small}
\begin{align}
\alpha & =-\frac{1}{2}arg\left(\frac{A_2 B_3 - A_3 B_2}{A_3 B_1 - A_1 B_3}\right), \\
\beta & =-\frac{1}{2}arg\left(\frac{A_3 B_2 - A_2 B_3 }{A_2 B_1- A_1 B_2}e^{2i\delta }\right).
\end{align}
\end{small}
Using the experimental inputs of the two mass-squared differences and the three mixing angles we can constrain the other parameters. Since $\Delta m_{21}^{2}$ and $|\Delta m_{23}^{2}|$ are known experimentally, the values of mass ratios $(\rho,\sigma)$ from Eqs. (13) and (14) can be used to calculate $m_1$.
This can be done by using Eq. (10) to obtain the two values of $m_1$ viz.,
\begin{small}
\begin{equation}
m_{1}=\sigma \sqrt{\frac{ \Delta
m_{21}^{2}}{1-\sigma ^{2}}} \ , \ \ 
m_{1}=\rho \sqrt{\frac{\Delta m_{21}^{2}+
|\Delta m_{32}^{2}|}{ 1-\rho^{2}}} \ .
\end{equation}
\end{small}
These two values of $m_1$ contain the constraints of two texture zeros in terms of $\rho~\textrm{and}~\sigma$ and must be equal to within the current experimental precision. 
There exists a permutation symmetry between different patterns \cite{xingzt2011} of two texture zeros. The corresponding permutation matrix has the following form: \begin{small} 
\begin{equation}
P_{23} = \left(
\begin{array}{ccc}
1&0&0\\
0&0&1\\
0&1&0\\
\end{array}
\right).
\end{equation}
\end{small}
The two texture zero neutrino mass matrices, therefore, are related to each other as
\begin{equation}
M_{\nu}^{X} = P_{23}M_{\nu}^{Y}P_{23}^T , 
\end{equation}
leading to the following relations between the neutrino oscillation parameters:
\begin{equation}
\theta_{12}^{X} = \theta_{12}^{Y}, \ \theta_{13}^{X} = \theta_{13}^{Y}, \ \theta_{23}^{X} = \frac{\pi}{2}-\theta_{23}^{Y}, \ \delta^{X} = \delta^{Y} - \pi \ 
\end{equation}
where X and Y denote the neutrino mass matrices related by the permutation symmetry $P_{23}$. The texture structures related by the 2-3 permutation symmetry in class B are
\begin{align}
B_1 \leftrightarrow B_2, \ B_3 \leftrightarrow B_4 \nonumber  
\end{align}

\subsection{Two Vanishing Cofactors in $M_\nu$}
The simultaneous existence of two vanishing cofactors in the neutrino mass matrix implies
\begin{eqnarray}
M_{\nu (pq)} M_{\nu (rs)}- M_{\nu
(tu)} M_{\nu (vw)}=0 \ , \\ M_{\nu (p'q')} M_{\nu (r's')}- M_{\nu
(t'u')} M_{\nu (v'w')}=0 \ .
\end{eqnarray}
These two conditions yield two complex equations viz.
\begin{eqnarray}
&&\sum_{k,l=1}^{3}(V_{pk}V_{qk}V_{rl}V_{sl}-V_{tk}V_{uk}V_{vl}V_{wl})m_{k}m_{l}=0 \ , \\
&&\sum_{k,l=1}^{3}(V_{p'k}V_{q'k}V_{r'l}V_{s'l}-V_{t'k}V_{u'k}V_{v'l}V_{w'l})m_{k}m_{l}=0 \ 
\end{eqnarray}
which can be rewritten as
\begin{eqnarray}
m_1 m_2 A_3e^{2i\alpha} + m_2 m_3 A_1e^{2i(\alpha+\beta +\delta )}+ m_3 m_1A_2e^{2i(\beta +\delta)}=0 \ , \\
m_1 m_2 B_3e^{2i\alpha} + m_2 m_3 B_1e^{2i(\alpha+\beta +\delta )}+ m_3 m_1 B_2e^{2i(\beta +\delta)}=0 \ ,
\end{eqnarray}
where
\begin{eqnarray}
A_h&=&(U_{pk}U_{qk}U_{rl}U_{sl}-U_{tk}U_{uk}U_{vl}U_{wl})+(k\leftrightarrow l) \ ,\\ \nonumber
B_h&=&(U_{p'k}U_{q'k}U_{r'l}U_{s'l}-U_{t'k}U_{u'k}U_{v'l}U_{w'l})+(k\leftrightarrow l) \ ,
\end{eqnarray}
with $(h,k,l)$ as the cyclic permutation of (1,2,3). Simultaneously solving Eqs. (25) and (26) for the two mass ratios, we obtain
\begin{small}
\begin{equation}
\frac{m_1}{m_2}e^{-2i\alpha }=\frac{A_3 B_1 - A_1 B_3}{A_2 B_3 - A_3 B_2}
\end{equation}
\end{small}
and
\begin{small}
\begin{equation}
\frac{m_1}{m_3}e^{-2i\beta }=\frac{A_2 B_1 - A_1 B_2 }{A_3 B_2 - A_2 B_3}e^{2i\delta } \ .
\end{equation}
\end{small}
The magnitudes of the two mass ratios in Eqs. (28) and (29) are given by
\begin{equation}
\rho=\left|\frac{m_1}{m_3}e^{-2i\beta }\right| ,
\end{equation}
\begin{equation}
\sigma=\left|\frac{m_1}{m_2}e^{-2i\alpha }\right| 
\end{equation}
 while the CP- violating Majorana phases $\alpha$ and $\beta$ are given by
 \begin{small}
\begin{equation}
\alpha =-\frac{1}{2}arg\left(\frac{A_3 B_1 - A_1 B_3}{A_2 B_3 - A_3 B_2}\right),
\end{equation}
\end{small}
\begin{small}
\begin{equation}
\beta =-\frac{1}{2}arg\left(\frac{A_2 B_1 - A_1 B_2 }{A_3 B_2 - A_2 B_3}e^{2i\delta }\right).
\end{equation}
\end{small}
Different classes of neutrino mass matrices with two vanishing cofactors are related by the permutation symmetry $P_{23}$ just as in case of neutrino mass matrices with two texture zeros. The texture structures related by 2-3 permutation symmetry in class B are
\begin{align}
B_3 \leftrightarrow B_4, B_5 \leftrightarrow B_6 \nonumber  
\end{align}

\section{Numerical Analysis}
The present experimental constraints on neutrino parameters at 1, 2 and 3$\sigma$ \cite{valledata} are given in Table 2.
For the numerical analysis, we generate $10^7$ random points ( $10^8$ when the number of allowed points is small).
\begin{table}[h]
\begin{center}
\begin{tabular}{cc}
\hline \hline Parameter & Mean$^{(+1 \sigma, +2 \sigma, +3 \sigma)}_{(-1 \sigma, -2 \sigma, -3 \sigma)}$ \\ \hline
$\Delta m_{21}^{2} [10^{-5}eV^{2}]$ & $7.62_{(-0.19,-0.35,-0.5)}^{(+0.19,+0.39,+0.58)}$ \\ 
 $\Delta m_{31}^{2} [10^{-3}eV^{2}]$ & $2.55_{(-0.09,-0.19,-0.24)}^{(+0.06,+0.13,+0.19)}$, \\&
$(-2.43_{(-0.07,-0.15,-0.21)}^{(+0.09,+0.19,+0.24)})$ \\ \hline
$\sin^2 \theta_{12}$ & $0.32_{(-0.017,-0.03,-0.05)}^{(+0.016,+0.03,+0.05)}$ \\ 
$\sin^2 \theta_{23}$ & $0.613_{(-0.04,-0.233,-0.25)}^{(+0.022,+0.047,+0.067)}$, \\& $(0.60_{(-0.031,-0.210,-0.230)}^{(+0.026,+0.05,+0.07)})$ \\ 
$\sin^2 \theta_{13}$ & $0.0246_{(-0.0029,-0.0054,-0.0084)}^{(+0.0028,+0.0056,+0.0076)}$,\\& $(0.0250_{(-0.0027,-0.005,-0.008)}^{(+0.0026,+0.005,+0.008)})$ \\ \hline \hline 
\end{tabular}
\caption{Current Neutrino oscillation parameters from global fits \cite{valledata}. The upper (lower) row corresponds to Normal (Inverted) Hierarchy, with $\Delta m^2_{31} > 0$ ($\Delta m^2_{31} < 0$).}
\end{center}
\end{table}
The effective Majorana mass of the electron neutrino $|M_{ee}|$ which determines the rate of neutrinoless double beta decay (NLDB) is given by
\begin{equation}
|M_{ee}|= |m_1c_{12}^2c_{13}^2+ m_2s_{12}^2c_{13}^2 e^{2i\alpha}+ m_3s_{13}^2e^{2i\beta}|.
\end{equation}
The observation of NLDB decay would signal lepton number violation indicating physics beyond the standard model. Its observation will imply the Majorana nature of neutrinos and, in addition, provide a way to probe the neutrino mass scale. There are a large number of projects such as CUORICINO\cite{cuoricino}, CUORE \cite{cuore}, MAJORANA \cite{majorana}, SuperNEMO \cite{supernemo}, EXO \cite{exo}, GENIUS \cite{genius} which aim to achieve a sensitivity upto 0.01 $eV$ for $|M_{ee}|$.  In our numerical analysis, we take the upper limit of $|M_{ee}|$ to be 0.5 $eV$.
In addition, the cosmological data put an upper limit on the sum of the active neutrino masses:
\begin{align}
\Sigma = \sum ^{3}_{i=1}m_{i}.
\end{align} 
The experimental results from Planck \cite{Planck} limit $\sum < 0.23~eV$ at $95\%$ confidence level (CL). However, these bounds strongly depend on the model details as well as on the data set used. In the present work, we take a conservative upper limit on $\sum$ to be 1 $eV$.
\begin{figure}[h]
\begin{center}
\epsfig{file=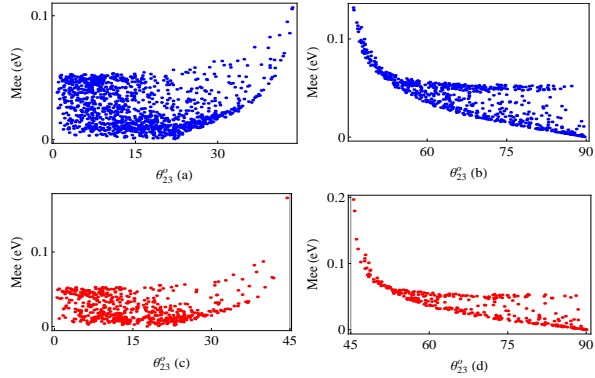,height=5.cm,width=7.8cm}
\end{center}
\caption{Correlation plots for $B_{3}$ of TZ(upper row) and TC(lower row) for NH(left panel) and IH(right panel) specturm.}
\end{figure}
\begin{figure}[!]
\begin{center}
\epsfig{file=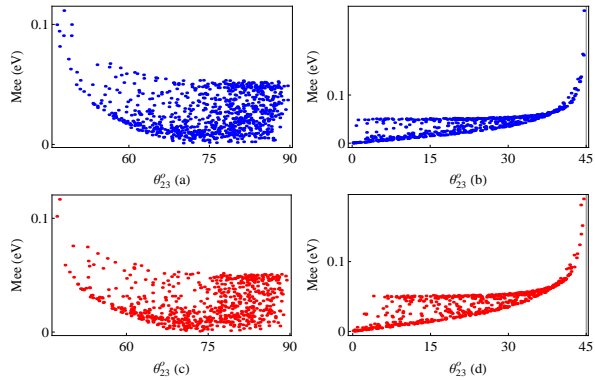,height=5.cm,width=7.8cm}
\end{center}
\caption{Correlation plots for $B_{4}$ of TZ(upper row) and TC(lower row) for NH(left panel) and IH(right panel) specturm.}
\end{figure}

In our numerical analysis, we vary the two mass-squared differences ($\Delta m^2_{21}$, $\Delta m^2_{31}$) randomly within their 3$\sigma$ allowed ranges but keep the other oscillation parameters free. The three neutrino mixing angles and Dirac-type CP-violating phase $\delta$ are varied between $0^{\circ}$-$90^{\circ}$ and  $0^{\circ}$-$360^{\circ}$, respectively. We perform the numerical analysis for class B of two texture zeros (TZ) and two vanishing cofactors(TC).
It is observed that the octant of $\theta_{23}$ is well restricted for $B_3$, $B_4$ of TZ and TC (see Figs.[1, 2]). This result is independent of the values of solar and reactor mixing angles and no constraint on $|M_{ee}|$ is required. However, for other classes of TZ and TC the value of $\theta_{23}$ is unconstrained like other oscillation parameters.

In the second step, we use the constraint of large $|M_{ee}|$ along with the experimental range of $\Delta m^2_{21}$, $\Delta m^2_{31}$ while keeping the mixing angles and the Dirac-type CP violating phase $\delta$ free. It is found that all the textures of class B of TZ as well as TC predict near maximal neutrino atmospheric mixing $\theta_{23}$. Using the experimental constraints on neutrino mixing  angles supplemented with a large $|M_{ee}|$, class B of TZ as well as TC predict near maximal atmospheric mixing angle $\theta_{23}$ and Dirac-type CP-violating phase $\delta$ fixed near $\pi/2$ or 3$\pi/2$.
\begin{figure}[h]
\begin{center}
\epsfig{file=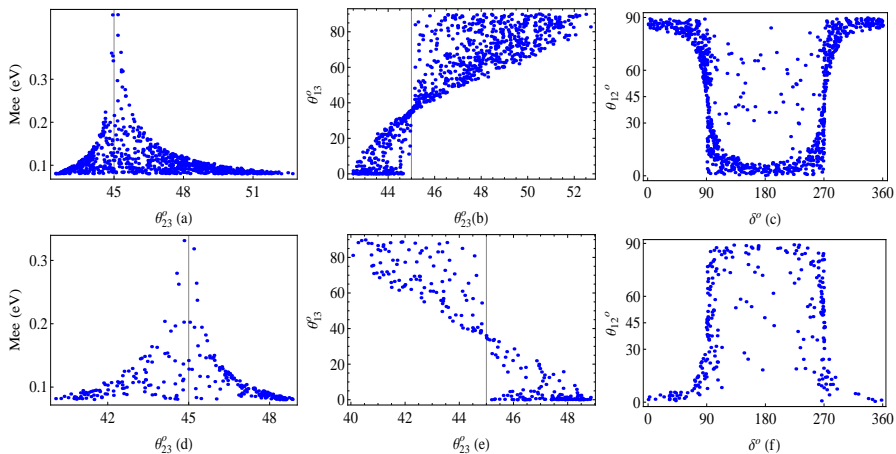,height=6.0cm,width=11.8cm}
\end{center}
\caption{Correlation plots for $B_{1}$ of TZ for NH(upper row) and IH(lower row) spectrum.}
\end{figure}
\begin{figure}[h]
\begin{center}
\epsfig{file=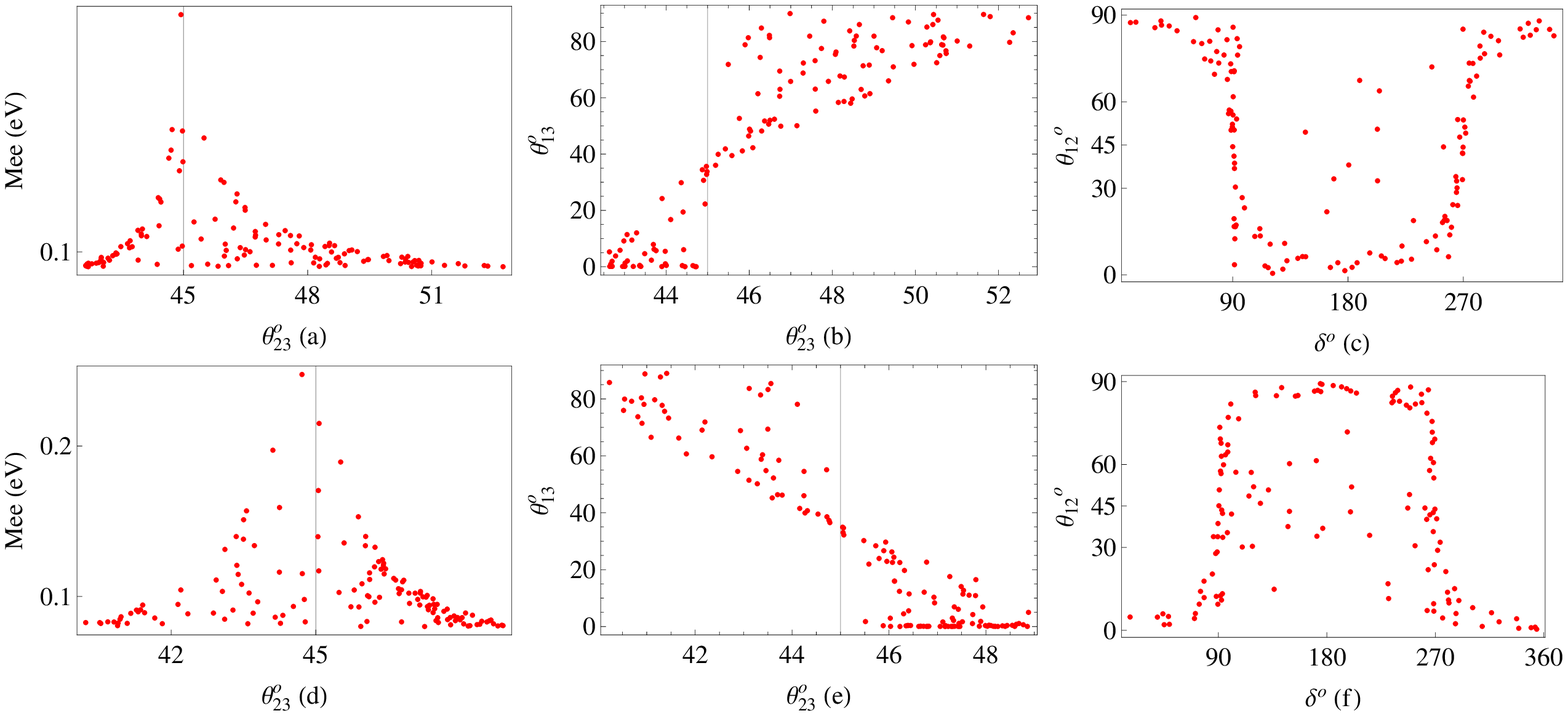,height=6.0cm,width=11.8cm}
\end{center}
\caption{Correlation plots for $B_{5}$ of TC for NH(upper row) and IH(lower row) spectrum.}
\end{figure}
These results are independent of the values of solar($\theta_{12}$) and reactor ($\theta_{13}$) neutrino mixing angles as noted by Dev et. al. in Ref. \cite{tzm} for textures $B_5$ and $B_6$ of TC and in Ref. \cite{tz_mee} for class B of TZ. The correlation plots for the oscillation parameters are shown in Fig.[3] and Fig.[4].
\begin{figure}[h]
\begin{center}
\epsfig{file=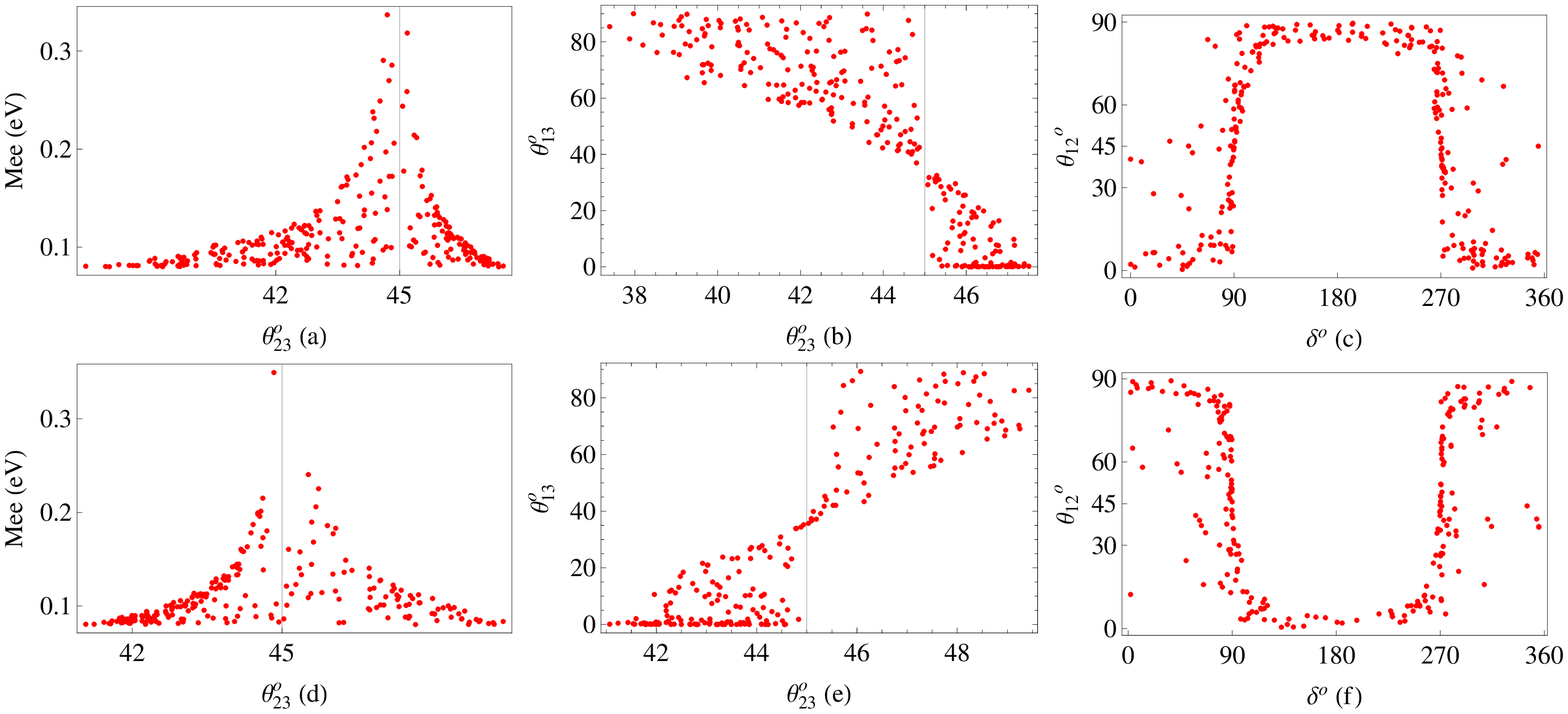,height=6.0cm,width=11.8cm}
\end{center}
\caption{Correlation plots for $B_{6}$ of TC for NH(upper row) and IH(lower row) spectrum.}
\end{figure}

It is evident from Fig.[3] and Fig.[4] that class $B_1$ of TZ and class $B_5$ of TC show similar predictions for the same mass hierarchy. It is clear  from Fig.[3] and Fig.[5] that classes $B_1$ of TZ and $B_6$ of TC with opposite mass hierarchy are distinguishable on the basis of the oscillation parameter $\delta$. This feature has not been discussed in Ref. \cite{Liao} while considering the indistinguishable nature of these texture structures. This distinction, in fact, becomes more clear if texture structures under investigation are subjected to the experimental inputs of neutrino mixing angles $\theta_{12}$, $\theta_{23}$ along with $\Delta m^2_{21}$, $\Delta m^2_{31}$ and large $|M_{ee}|$ while allowing $\theta_{13}$ to vary between $0^{\circ}$-$12^{\circ}$. All these texture structures show a marked preference for a particular quadrant of $\delta$ as given in Table 3. It is clear from the Table 3 that a distinction between $B_1$($B_2$) of TZ and $B_6$($B_5$) of TC can be made on the basis of quadrant of CP violating phase $\delta$ even if neutrino mass hierarchy is not known.
\begin{table}[h]
\begin{small}
\begin{center}
\begin{tabular}{cccc}\hline \hline
 Class  & $\delta$ quadrant & Normal Hierarchy  & Inverted Hierarchy \\ 
\hline 
&&Two Texture Zeros \\ \hline
 $B_1$ & (2,3) & $\theta_{23}<\pi/4$ & $\theta_{23}>\pi/4$ \\ 
 $B_2$ & (1,4) & $\theta_{23}>\pi/4$ & $\theta_{23}<\pi/4$\\ 
 $B_3$ & (1,4) & $\theta_{23}>\pi/4$ & $\theta_{23}<\pi/4$ \\ 
 $B_4$ & (2,3) & $\theta_{23}<\pi/4$ & $\theta_{23}>\pi/4$\\ 
 \hline \hline
 &&Two Vanishing Cofactors \\ \hline
 $B_3$ & (1,4) & $\theta_{23}<\pi/4$ & $\theta_{23}>\pi/4$ \\ 
 $B_4$ & (2,3) & $\theta_{23}>\pi/4$ & $\theta_{23}<\pi/4$\\ 
 $B_5$ & (2,3) & $\theta_{23}<\pi/4$ & $\theta_{23}>\pi/4$ \\ 
 $B_6$ & (1,4) & $\theta_{23}>\pi/4$ & $\theta_{23}<\pi/4$\\ 
 \hline \hline
\end{tabular}
\caption{Experimental signatures distinguishing class B of two texture zeros \cite{tz}(two vanishing cofactors\cite{Lashin}).}
\end{center}
\end{small} 
\end{table}
It is important to keep in mind that the results obtained in Table 3 are for the values of $\theta_{13}$ between  $0^{\circ}$ to $12^{\circ}$ as the exact value of $\theta_{13}$ was not known at that time. However, a change in the quadrant of $\delta$ which is in marked contrast from the results tabulated in Table 3 has been predicted for a relatively large value of $\theta_{13}$ presently known. Interestingly, this change has been predicted only for the inverted mass hierarchy while the results for the normal hierarchy spectrum are in conformity with the results from earlier analysis presented in Table 3.
\begin{figure}[!h]
\begin{center}
\epsfig{file=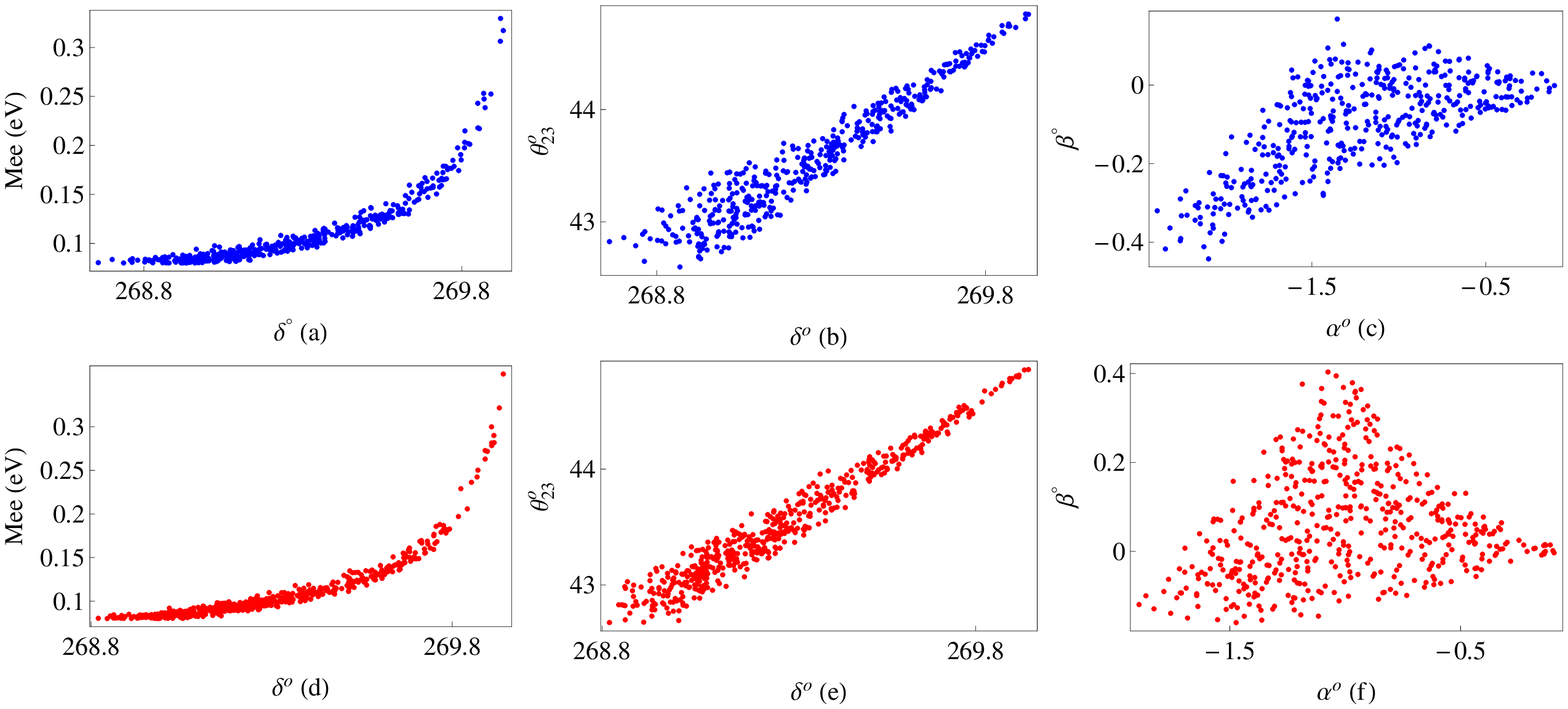,height=6.0cm,width=11.8cm}
\end{center}
\caption{Correlation plots for $B_{1}$ of TZ(upper row) and $B_{5}$ of TC(lower row) for NH spectrum.}
\end{figure}
\begin{figure}[!h]
\begin{center}
\epsfig{file=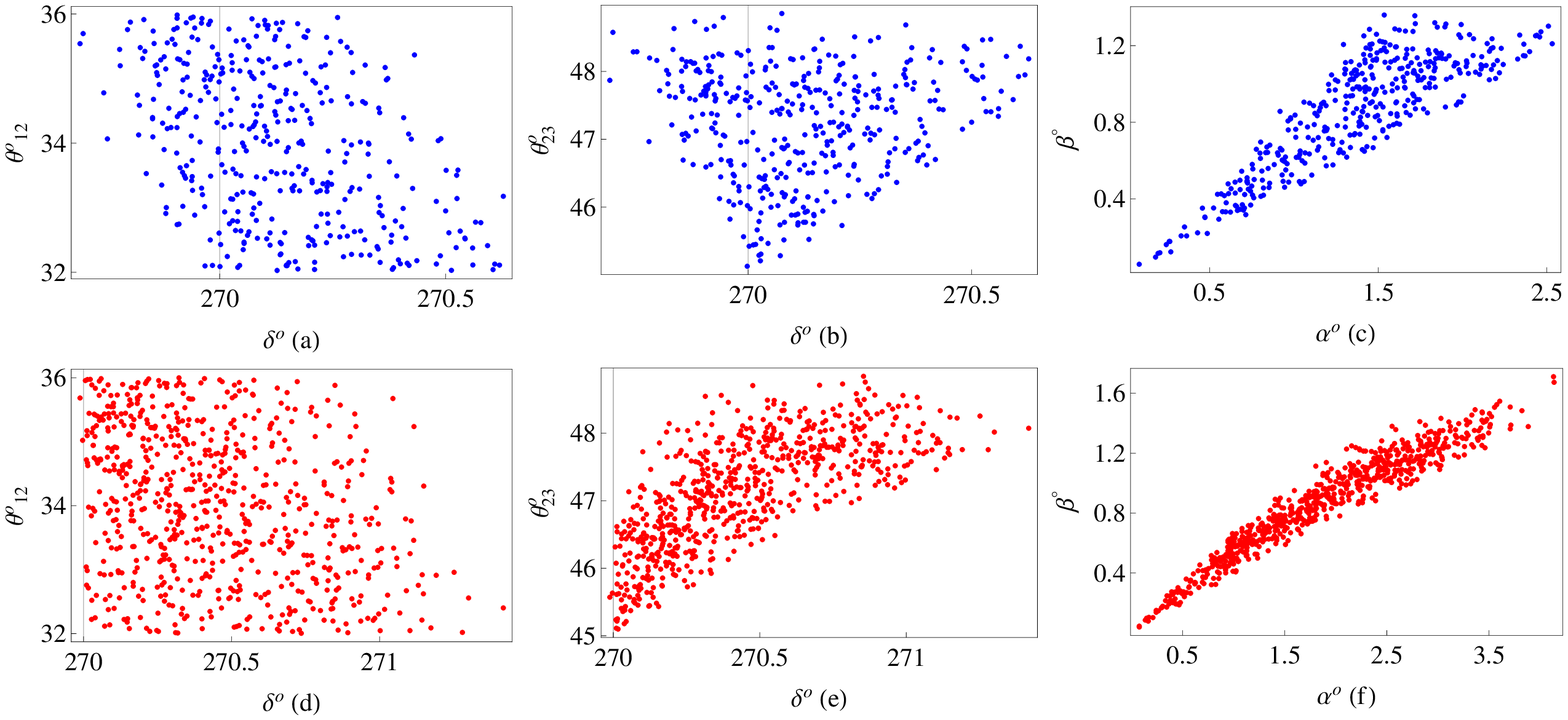,height=6.0cm,width=11.8cm}
\end{center}
\caption{Correlation plots for $B_{1}$ of TZ(upper row) and $B_{5}$ of TC(lower row) for IH spectrum.}
\end{figure}
\begin{figure}[h]
\begin{center}
\epsfig{file=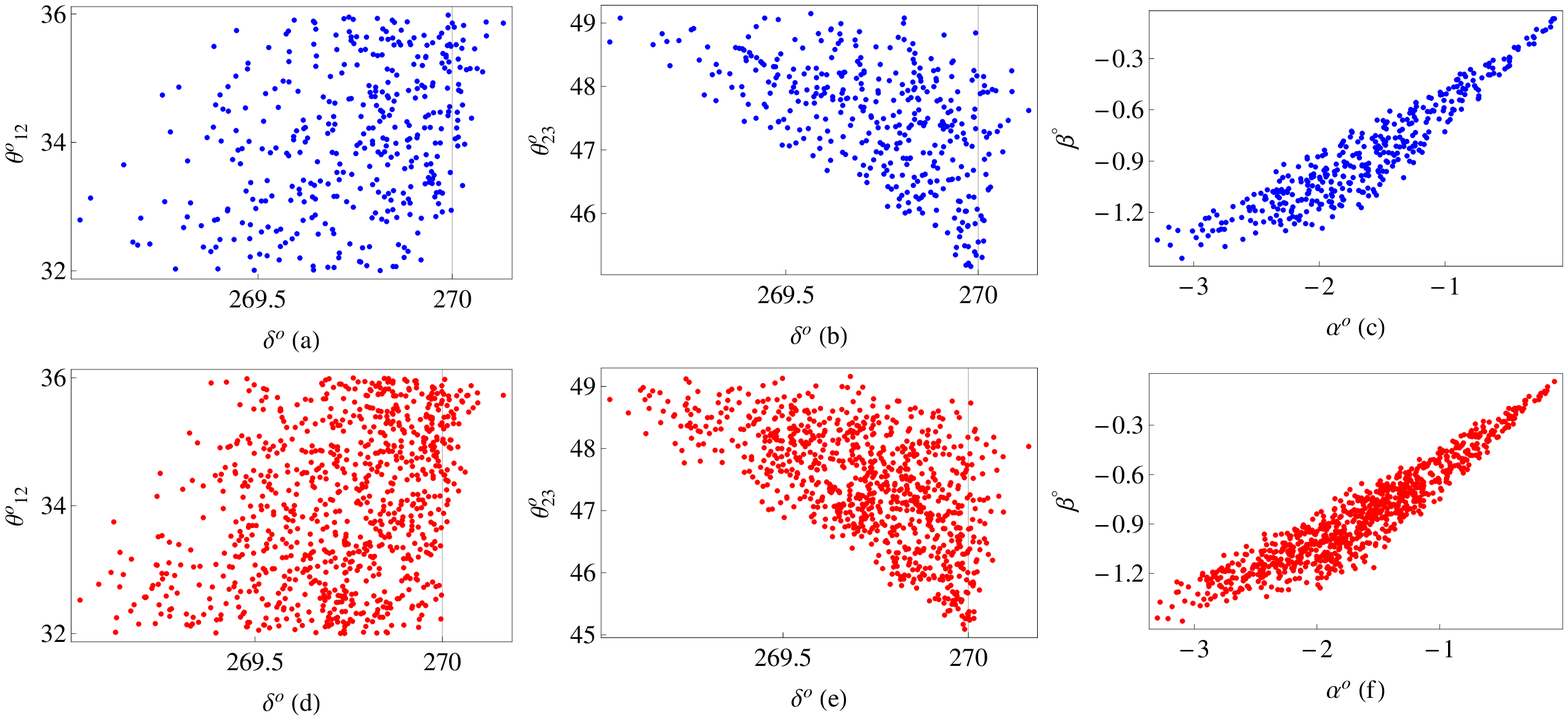,height=6.0cm,width=11.8cm}
\end{center}
\caption{Correlation plots for $B_{3}$ of TZ(upper row) and TC(lower row) for IH spectrum.}
\end{figure}

In the next step, we vary  $\delta$ randomly between $265^{\circ}$-$275^{\circ}$ and all other parameters within their 3$\sigma$ CL ranges to explore the quadrant of $\delta$ more accurately in the light of the current experimental data. Fig.[6] shows correlation in $B_1$ of TZ and $B_5$ of TC for normal mass spectrum and Fig.[7] shows correlation in $B_1$ of TZ and $B_5$ of TC for an inverted mass spectrum. It is evident from the plots that classes $B_1$($B_3$) of TZ and $B_{5}$($B_3$) of TC show nearly similar predictions for oscillation parameters. However, Fig.[6] reveals that there is a small difference in the Majorana phase $\beta$ for $B_1$ of TZ and $B_5$ of TC. Such differences can be there for $B_1$($B_2$) of TZ and $B_5$($B_6$) of TC, respectively as these textures are not related trivially as is the case for $B_3$($B_4$) of TZ and $B_3$($B_4$) of TC respectively. Distinguishing these texture structures depends on how precisely the Majorana phases are measured in the near future\cite{mphases}.
\begin{figure}[!]
\begin{center}
\epsfig{file=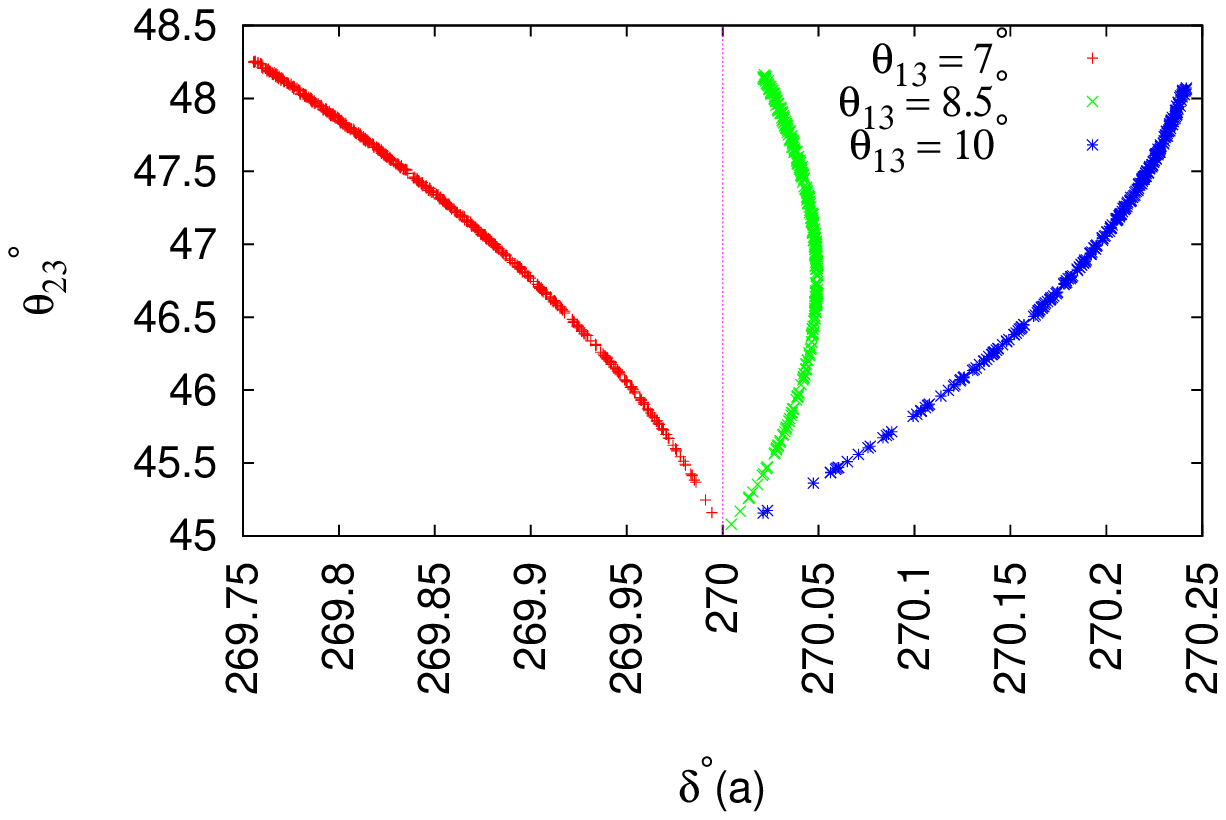,height=4.1cm,width=5.0cm}
\epsfig{file=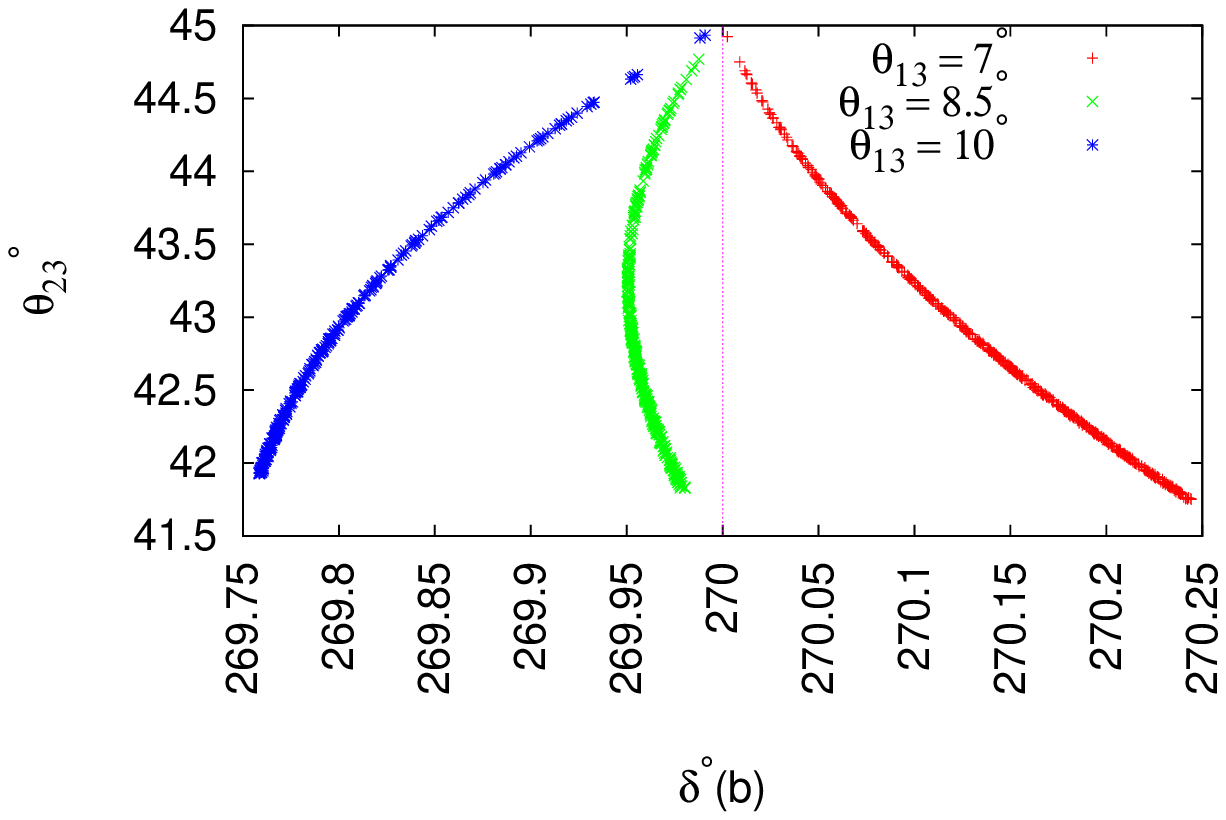,height=4.1cm,width=5.0cm}\\
\epsfig{file=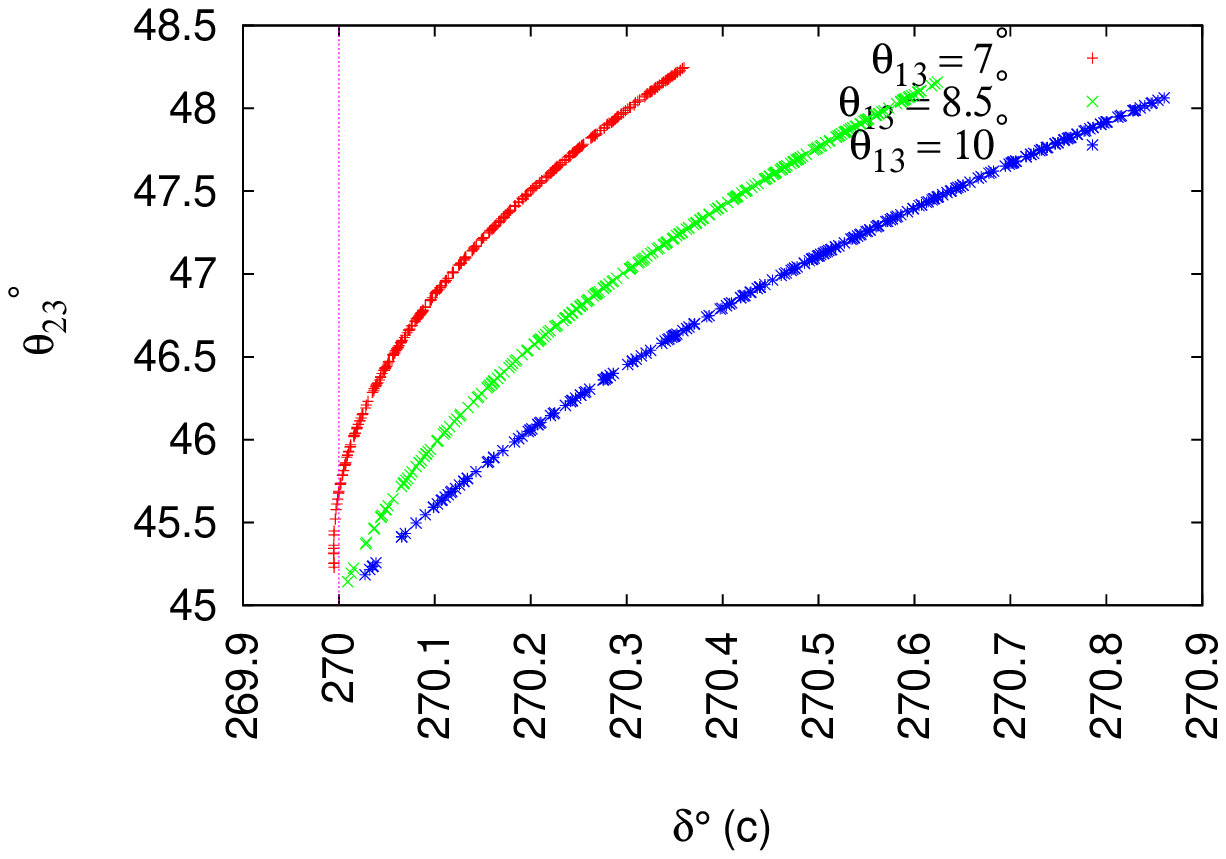,height=4.1cm,width=5.0cm}
\epsfig{file=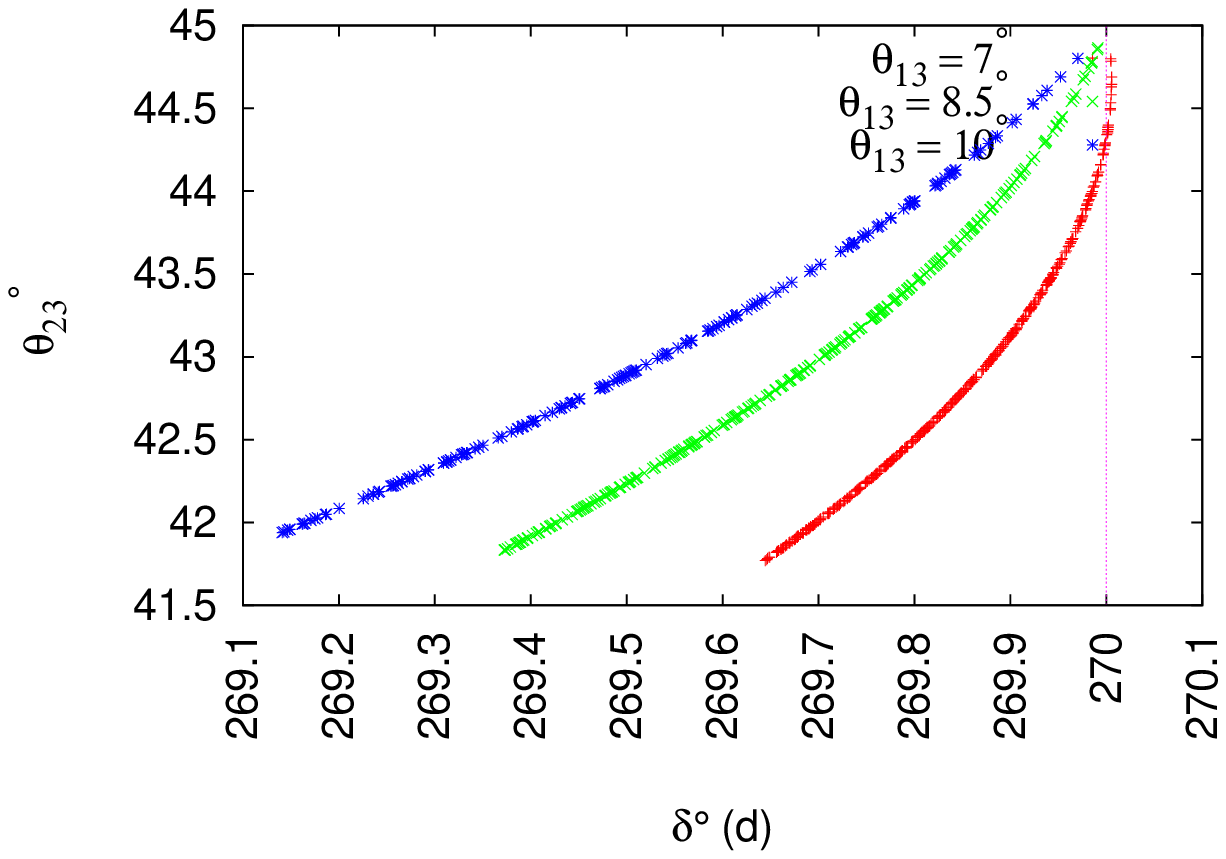,height=4.1cm,width=5.0cm}
\end{center}
\caption{Correlation between $\theta_{23}$ and $\delta$ for $B_{1}(a)~\textrm{and}~B_{2}(b)$ of TZ (upper row) and $B_{5}(c)~\textrm{and}~B_{6}(d)$ of TC (lower row) for IH spectrum.}
\end{figure}
\begin{figure}[!]
\begin{center}
\epsfig{file=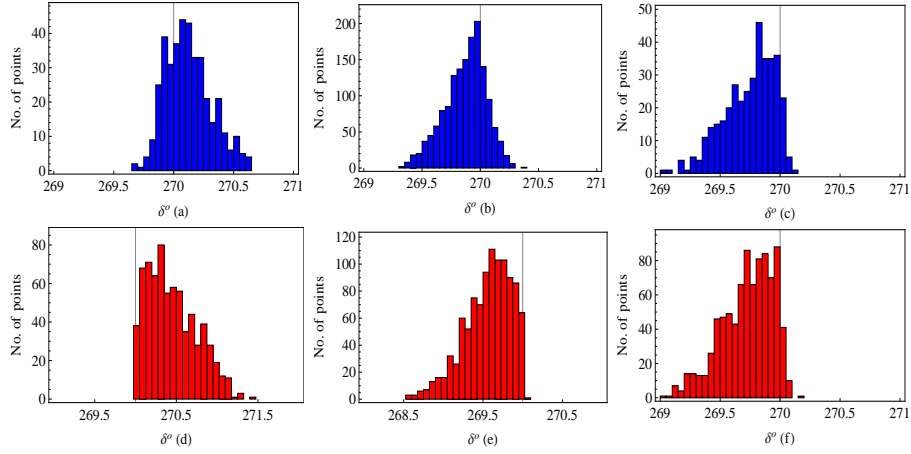,height=6.0cm,width=12.0cm}
\end{center}
\caption{Histogram plots for $B_1$(a), $B_2$(b), $B_3$(c) of TZ and $B_5$(d), $B_6$(e), $B_3$(f) of TC for IH spectrum.}
\end{figure}

Fig.[8] shows correlation in textures $B_3$ of TZ and TC for IH spectrum. The textures $B_3$ of TZ and TC show similar predictions for oscillation parameters which is expected as these texture structures are trivially related to each other. The same is the case for textures $B_{4}$ of TZ and TC. Correlation plots in Fig.[1] and Fig.[2] depict the permutation symmetry between $B_3$ and $B_4$ class of TZ and TC and, also, the  permutation symmetry between $B_5$ and $B_6$ texture structures of TC is evident from Fig.[4] and Fig.[5]. The results for the 2-3 interchange symmetric textures which are not discussed here can simply be obtained using Eq.  (20). Fig.[9] shows the correlation plots for $B_1$, $B_2$ of TZ and $B_5$, $B_6$ of TC for IH spectrum. The plots show correlation between $\theta_{23}$ and $\delta$ for  different values of $\theta_{13}$ while all other parameters are at their best fit values. It is clear from the figure  that there is a preference for change in the quadrant of $\delta$ in contrast to the results tabulated in Table 3 for the presently known relatively large value of reactor mixing angle $\theta_{13}$ for IH spectrum. For relatively large value of $\theta_{13}$, the Dirac-type CP-violating phase $\delta$ shows a preference for the (1,4) quadrant in class $B_{1}$($B_5$) of TZ(TC) and a preference for the (2,3) quadrant in class $B_3$ of TZ and TC in contrast to the results summarized in Table 3. However, for normal hierarchical mass spectrum the marked preference for a particular quadrant of $\delta$ in class B of TZ and TC is still there and is the same as given in Table 3. To see the preference for shift in quadrant of $\delta$ more clearly for inverted hierarchical spectrum, histograms have been plotted for class B of TZ and TC(see Fig.[10]). The histograms clearly show the most probable values for Dirac-type CP violating phase $\delta$ varied between the range $265^{\circ}$-$275^{\circ}$ whereas all other neutrino oscillation parameters are varied within their  experimental $3\sigma$ ranges. One can see from the histograms that all the texture structures of class B show a preference for shift in the quadrant of $\delta$ in contrast to results of the earlier analyses given in Table 3 for TZ and TC for the present neutrino oscillation data with relatively large $\theta_{13}$. 

\section{Summary}
In the present work, phenomenological analysis for classes B of TZ and TC has been presented in the light of the recent neutrino oscillation data. It has been noted that the quadrant of $\theta_{23}$ is independent of the values of $\theta_{12}$ and $\theta_{13}$ and does not require the constraint of large value of $|M_{ee}|$ for $B_3$, $B_{4}$ textures of TZ and TC. Classes $B_1$($B_2$) of TZ and $B_5$($B_6$) of TC have been found to have similar predictions for similar mass hierarchy except for a small difference in Majorana phases which can be due to the non-trivial structures of TZ and TC for these texture structures. The preference for a particular quadrant of the CP violating phase $\delta$ in contrast to the results reported in earlier analyses has been predicted for class B of TZ and TC for an inverted mass spectrum. 
 
\textbf{\Large{Acknowledgements}}\\
L. S. and D. R. gratefully acknowledge the financial support provided by University Grants Commission (UGC), Government of  India.

\end{document}